\documentclass[galaxies,review,accept,pdftex,oneauthor]{Definitions/mdpi} 

\firstpage{1} 
\pubvolume{13}
\issuenum{1}
\articlenumber{66}
\pubyear{2025}
\copyrightyear{2025}
\datereceived{4 April 2025} 
\daterevised{27 May 2025} 
\dateaccepted{30 May 2025} 
\datepublished{5 June 2025} 
\hreflink{https://doi.org/10.3390/
galaxies13030066} 

\Title{Red Supergiants in the Milky Way and Nearby Galaxies}

\TitleCitation{Red Supergiants in the Milky Way and Nearby Galaxies}

\Author{Alceste Z. Bonanos~\orcidA{}}

\AuthorCitation{Bonanos, A.Z.}

\address[1]{Institute for Astronomy, Astrophysics, Space Applications and Remote Sensing, 
 National Observatory of Athens, GR-15326 Penteli, Greece; bonanos@astro.noa.gr}

\abstract{Identifications of red supergiants (RSGs) in the Milky Way and nearby galaxies have experienced an exponential increase in recent years, driven by advancements in selection techniques, the continued expansion of archival datasets, and a steady increase in spectroscopic data. This review describes the advances in methodologies and selection criteria for identifying RSGs and presents the current census of these stars in our own Galaxy and nearby galaxies. It also describes the insights gained from resolving nearby RSGs and their complex circumstellar material in the Milky Way and from the growing samples of RSGs being discovered in the Local Group and beyond. These advances impact the Humphreys--Davidson limit in the cool part of the Hertzsprung--Russell diagram. Furthermore, they provide insight into extreme RSGs and the role of photometric variability and, in particular, of the newly discovered phenomenon of dimming events. Recent observations have enabled the determination of the binarity fraction among RSGs, offering new constraints to stellar evolution. Looking ahead, the synergy between large-scale surveys, high-resolution observations, and emerging machine-learning tools promises to further transform our understanding of the final evolutionary stages of massive stars in the \mbox{coming decade.}}


\begin{document}


\section{Introduction}
Over the last decade, there has been significant progress enabling in-depth studies of individual RSGs and the characterization of entire RSG populations in nearby galaxies. Improved instrumentation has led to the availability of multi-band photometric catalogs of nearby galaxies and time-series photometry of thousands of stars in the Galaxy, reaching out to several Mpc, while multi-object spectroscopy has enabled detailed studies of hundreds of RSGs in both the optical and near-infrared. This has allowed for a comprehensive view of RSGs, ranging from the lowest luminosities ($\log L/L_{\odot} \sim 4$) to the Humphreys--Davidson limit \citep{Humphreys1979} at $\log L/L_{\odot} \sim 5.8$. We have entered the era of large RSG samples, which enable us to tackle open questions regarding their evolution, wind driving mechanism, mass-loss rates, and variability. We can also resolve RSGs and learn about their mass-loss history. The following sections review the improvements and limitations of the selection criteria and recent studies of the RSG population of the Milky Way, the Local Group, and beyond. The final sections present insights gained from extragalactic RSGs and a future outlook.


\section{Selection Criteria}

The first step in investigating RSGs requires robust selection criteria to identify the objects of interest and distinguish them from contaminants, such as red dwarfs, red giants, {distant red galaxies, or other reddened objects}. The initial studies, dating back to the 1970s, confirmed RSGs by obtaining long-slit spectra of candidates (resulting from objective-prism studies or color criteria) in the region of the CaII triplet, e.g., \citep{Humphreys1970}, or in the blue, e.g., \citep{Humphreys1979b}. Significant progress has been made over the last three decades in improving the selection criteria and making them broadly applicable, although spectra are still required to confirm the classification. 

In 1998, \citet{Massey1998a} discovered that low surface gravity red stars separate from red dwarfs on a $B-V$ vs. $V-R$ color diagram, providing the first selection criteria to identify the RSG population in nearby galaxies. This discovery stemmed from the multi-band survey of Local Group galaxies undertaken by Massey and collaborators in 1999--2001 using the Curtis Schmidt telescope for the Magellanic Clouds \citep{Massey2002} (with its wide-field CCDs) and the KPNO and CTIO 4 m telescopes for more distant Local Group galaxies \citep{Massey2007}. However, foreground stars remained a contaminant and spectroscopy was considered and still remains the most reliable way to confirm an RSG candidate. Furthermore, red giant and intermediate-mass AGB stars appear similar spectroscopically, see, e.g., 
 \citep{Brunish1986, Rayner2009}; therefore, distinguishing RSGs from these types of stars is challenging. \citet{Massey2003} were the first to use multi-object spectroscopy to study a population of over 100 RSGs in the LMC and SMC, finding a contamination fraction by foreground red dwarfs of 5\% and 11\%, respectively. Subsequent surveys of the Magellanic Clouds M31 and M33 (see Section~\ref{Section4}) obtained spectroscopy (in particular of the CaII triplet region) and used radial velocities to distinguish red foreground stars from RSGs in the target galaxies. 

\begin{figure}[H]
\begin{adjustwidth}{-\extralength}{-4cm}
\centering
\includegraphics[angle=270,width=0.6\columnwidth]{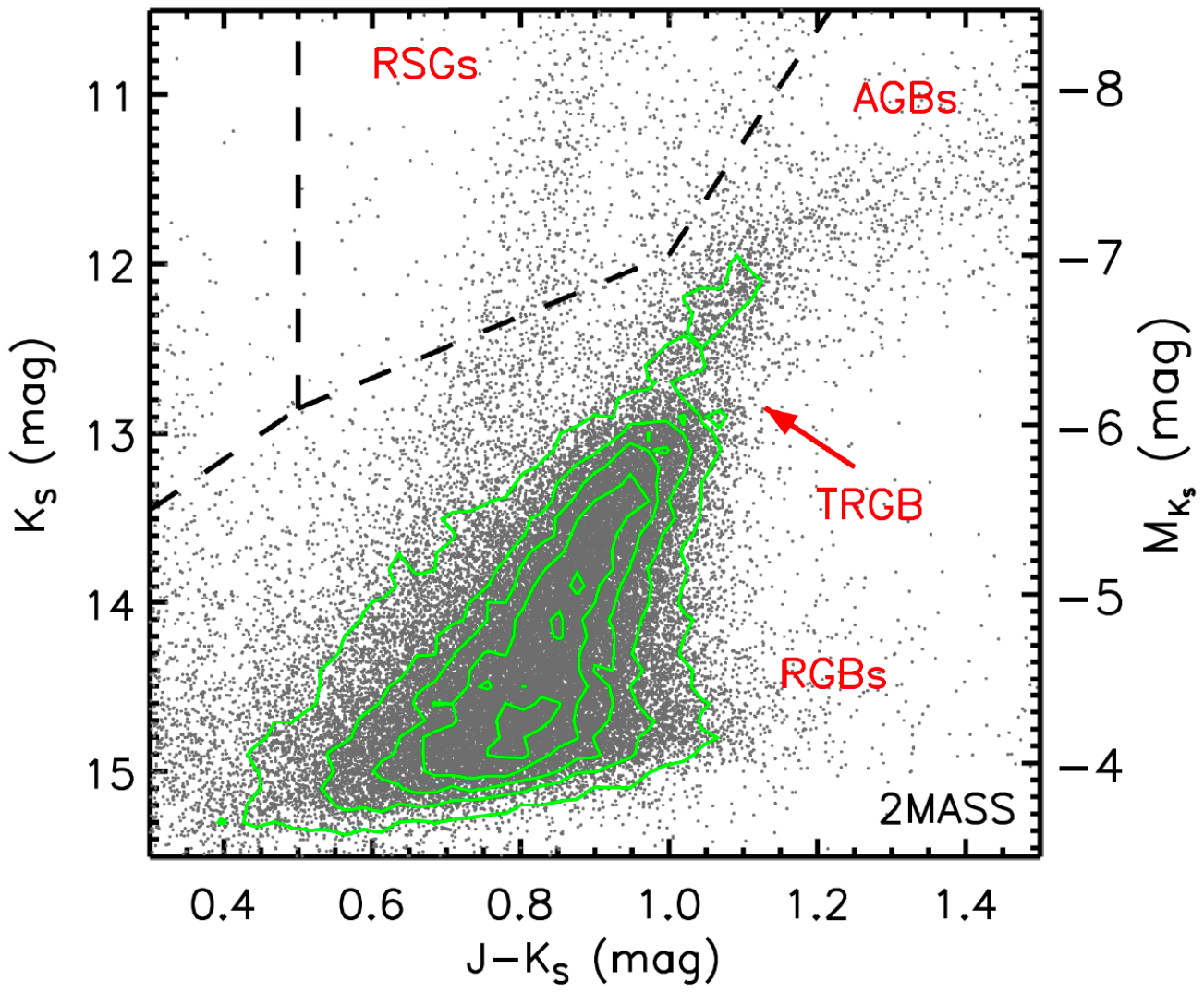}
\end{adjustwidth}
\caption{Separation of RSG from AGB in $K_s$ vs. $J-K_s$ CMD {in the SMC} (reproduced from \citep{Yang2019}). The dashed lines indicate the proposed criteria for distinguishing RSGs.}\label{fig1}
\end{figure}

The next breakthrough came with the 2MASS all-sky survey. As the RSG spectral energy distribution peaks in the near-IR, the IR regime was the next obvious region to develop selection criteria in. The infrared reddening-free parameter \mbox{$Q = (J-H)-1.8 \times (H-K_s)$} (see, e.g., \citep{Negueruela2007}) was used to select RSGs as stars with $Q$ = 0.1--0.4, 
 primarily in the Milky Way (see, e.g., \citep{Messineo2016}). Furthermore, color-cuts in color--magnitude and color--color diagrams using $V-K$, $J-H$, and $H-K$ were developed to identify RSG candidates and distinguish them from red giants and asymptotic giant branch (AGB) stars. \citet{Yang2019} demonstrated that AGB stars can be distinguished from RSG using near-IR colors, while \citet{Yang2021} demonstrated the separation of low and high surface gravity stars using the $H$-bump at 1.6~$\upmu$m. Moreover, the metallicity dependence of the optical and near-IR colors of RSGs has been explored in \citet{Li2025}. In Figure~\ref{fig1}, the RSG branch is distinct from the AGB branch and seems to extend to luminosities below the nominal cut-off at M$_{Ks}=-7$~mag, indicating a sequence of red He-burning stars, which extends below the ``massive star'' definition to intermediate mass stars. This raises the question of what the definition of a RSG should be and whether or not the requirement of core-collapse should be included. \citet{Yang2024} investigated the lowest 
 luminosity and mass limit of RSGs, finding it to extend down to $\log L/L_{\odot} \sim 3.5$ and an initial mass $\sim$6~M$_{\odot}$.

\begin{figure}[H]
\includegraphics[width=1\textwidth]{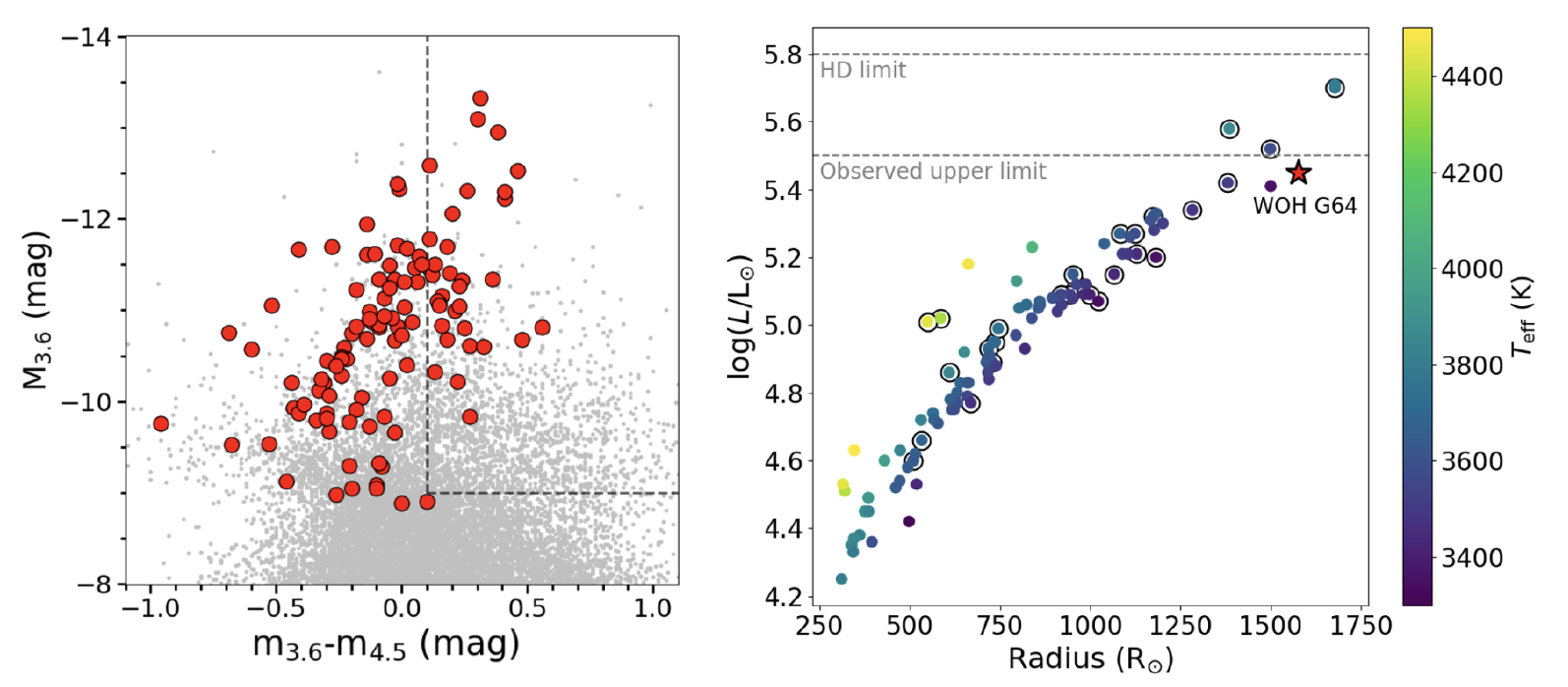}
\caption{(\textbf{Left}): [3.6]–[4.5] vs. M$_{[3.6]}$ color--magnitude diagrams of 129 RSGs in nearby galaxies from the ASSESS project \citep{Bonanos2024}. Gray background points are sources from the \textit{Spitzer} point source catalog of NGC 300. The dashed lines indicate the region of prioritized sources exhibiting IR excess, a sign of episodic mass loss. (\textbf{Right}): Stellar 
 radius versus luminosity for 92 of those RSGs that have a luminosity determination. Circles indicate
sources from the dusty sample, four of which (including three dusty
ones) exceed the {observed upper limit}. The extreme RSG WOH~G64 is marked by a red star. (reproduced from \citep{deWit2024}).}\label{fig2}
\end{figure}

{The advent of \textit{Spitzer} 
 and its Legacy surveys of the Magellanic Clouds \citep{Meixner2006, Gordon2011} lead to mid-IR ``roadmaps'' for interpreting luminous massive stars \citep{Bonanos2009, Bonanos2010}. RSGs were found to be among the brightest mid-IR sources, due to their intrinsic brightness and due to being surrounded by their own dust. This enabled the development of new mid-IR criteria (e.g., [3.6]--[4.5] 
 and $J-[3.6]$ vs. [3.6]) to identify RSGs and in particular dusty RSGs (with [3.6]--[4.5] $>$ 0.1 mag), as circumstellar dust surrounding RSGs produces excess infrared emissions. The fact that the majority of RSGs have mid-IR colors [3.6]--[4.5] $\leq$ 0 (see Figure}~\ref{fig2}) are likely due to the presence of CO and SiO bands in these stars affecting the {[4.5] band}~\citep{Verhoelst2009}, while RSGs with high mass-loss rates exhibit increased crystalline silicates~\citep{Waters2010}. The \textit{WISE} and NEOWISE missions have also produced mid-IR data for the whole sky; however, their resolution limits the application of the mid-IR criteria with \textit{WISE} to our Milky Way. A combination of near- and mid-IR criteria was suggested \mbox{by \citet{Messineo2012}} for the Galactic plane. The mid-IR criteria have proven successful in selecting dusty massive stars~\citep{Britavskiy2015}. However, they picked up HII regions, LBVs, sgB[e] stars, and emission objects as contaminants. The methodology was improved by the ASSESS project~\citep{Bonanos2024}, which also considered the optical counterparts, as well as a visual check to reject extended sources (e.g., galaxies) using archival images, yielding a success rate of 36\% in selecting evolved massive stars. Furthermore, ASSESS demonstrated that RSGs with IR excess [3.6]--[4.5] $>$ 0.1~mag have undergone episodic mass loss. Currently, JWST, which can only survey portions of nearby galaxies given its small field of view, is providing very deep, high-resolution images and accurate near- and mid-IR photometry for galaxies at $\sim$10~Mpc (e.g., PHANGS-JWST \citep{Lee2023, Williams2024}), at a similar quality as the performance of \textit{Spitzer} at 1~Mpc, since the angular resolution of JWST is $\sim$10 times better. \citet{Levesque2018} have investigated near-IR diagnostics for selecting RSGs from foreground sources, while \citet{Boyer2024} investigated color--color diagrams with JWST filters in the range between 0.9 and 4.3 $\upmu$m to separate RSGs from AGB stars in the {Wolf–Lundmark–Melotte (WLM) galaxy}. Spiral galaxies beyond about 10 Mpc can fit in the field of view of JWST; therefore, JWST is well suited to detect and characterize entire populations of RSGs in a variety of environments and metallicities. 

\vspace{-6pt}
\unskip

The \textit{Gaia} mission has measured proper motions and parallaxes for billions of stars (in data releases DR2 and DR3 \citep{Gaia2016, gaia2018, Gaia2023}), providing a way to eliminate or at least greatly reduce foreground red giants, which plagued previous methodologies, and have lead to a great improvement in the selection criteria. \citet{Aadland2018} were the first to apply \textit{Gaia} parallaxes to eliminate foreground stars in the direction of the LMC. \mbox{\citet{Yang2019}} and \citet{Maravelias2022} similarly performed \textit{Gaia} cleaning to eliminate foreground stars toward the SMC and M31 and M33, respectively, by modeling the proper motion distribution of the foreground vs. host galaxy stars. This method was applied in \mbox{\citet{Bonanos2024}} and \mbox{\citet{deWit2025}} 
 to select dusty RSGs in nearby galaxies and helped reduce the foreground contamination (to 13\%). \textit{Gaia} also provided low-resolution spectra, which have been exploited to determine classification criteria based on the spectral features near the calcium triplet \citep{Dorda2016}, to search for RSGs \citep{Messineo2023, Zhang2024}, and as a training set for deriving the parameters of RSGs with machine learning \citep{DornWallenstein2023}.

During the last few years, machine-learning techniques have emerged as a very promising method to identify RSGs in nearby galaxies to distances beyond 10~Mpc. \citet{DornWallenstein2021} and \citet{Maravelias2022} were the first to apply machine-learning techniques and develop algorithms to classify evolved massive stars. These two independent studies were both based on optical and infrared color indexes (the former also included photometric variability from the infrared light curves), and both found their algorithms to have a success rate of recovering RSGs that is over $90\%$. \citet{Maravelias2025} has applied the algorithm to over 1 million sources in 26 nearby galaxies within 5~Mpc, producing accurate classifications for over 275,000 sources. These include over 120,000 RSGs and likely contain the entire RSG population of M31, M33, NGC 6822, and IC 1613. As the input data improve with the increasing amount of JWST data of nearby galaxies becoming available, and soon data from Rubin LSST, these algorithms will be able to achieve even higher success in identifying RSGs.

An underexploited tool for selecting RSGs is time-series photometry, as RSGs exhibit semi-regular or irregular variability, with a typical amplitude of $\sim$1~mag in the optical, and in extreme cases up to 4 mag \citep{Kiss2006, Szczygiel2010}, while in the mid-infrared the amplitudes are typically up to 0.5 mag \citep{Yang2018}. Variability in RSGs is due to both the pulsations and motions of large convective cells, while the extreme cases suggest a superwind phase. The amplitude of optical variability correlates with luminosity, e.g., \citep{Soraisam2018}; moreover, a period--luminosity relation has been established in various bands \citep{Yang2012, Ren2019, Chatys2019}, with the $K$-band having the least scatter. Granulation, which appears on the surface of RSGs due to the motion of convective cells and manifests as irregular variability, was presented as a new method for identifying and probing the properties of RSGs by \citet{Ren2020}. They analyzed the light curves of a large sample of RSGs and found characteristic amplitudes of 10--1000~mmag and evolution timescales of the granulations between several days to 1~yr. These quantities were also shown to increase with metallicity and to correlate with the physical properties of the RSGs, demonstrating the importance of variability as a tool to study RSGs.  The recent availability of optical data from transient surveys such as PanSTARRS1, ATLAS, ZTF, and \textit{Gaia} (and in particular, \textit{Gaia} DR4 expected in 2026) and mid-infrared light curves from AllWISE and NEOWISE allow variability information to be considered in a systematic way in the identification of RSGs. Within the next decade, high-cadence observations will become available for thousands of RSGs in nearby galaxies, launching a time-domain revolution in studies of evolved massive stars, and in particular, RSGs.

Once identified, optical and near-IR RSG spectra can be obtained and classified using the atlases of \citet{Rayner2009} or \citet{Messineo2021}, and physical parameters can be extracted. {This is currently feasible only for RSGs within a few Mpc, given the limiting magnitude achievable with existing instrumentation on 10 m class telescopes.} The effective temperature, T$_{\rm eff}$, can be derived from the spectra using 1D MARCS models \citep{Gustafsson2008}, corrected for non-local thermal equilibrium effects \citep{Bergemann2012, Bergemann2013, Bergemann2015}, or alternatively by using empirical or synthetic relations based on $J-K_s$ colors, e.g., \citep{deWit2024}. The luminosities (L) can be obtained via SED fitting (or metallicity-dependent bolometric corrections in the $K$-band, BC$_K$). The radii (R) can then be derived from the Stefan--Boltzmann law.

\section{RSG Population of the Milky Way}

The RSG population of the Milky Way remains largely unexplored. This is because the discovery and characterization of RSGs in our Galaxy faces two obstacles: high extinction values and uncertain distances. RSGs are mainly found in OB associations and clusters that are located in the Galactic disk; therefore, extinction becomes an increasing problem with distance. The case of Betelgeuse illustrates the distance issue: despite being the brightest, most well-known and possibly the most well-studied (with $>$150 refereed publications including `Betelgeuse' in the title) 
 RSG in the sky, its uncertain distance {and wavelength-dependent angular diameter, e.g., \citep{Haubois2009, Neilson2011, Cannon2023}, affect} the interpretation of the observations. Current distance estimates range from 168$^{+27}_{-15}$ pc (based on a seismic analysis \citep{Joyce2020}) to 222$^{+48}_{-34}$ pc (based on an updated astrometric solution \citep{Harper2017}). This distance range translates to a range in radii between 750 and 1000~R$_{\odot}$. 

\begin{figure}[H]
\begin{adjustwidth}{-\extralength}{-4cm}
\centering
\includegraphics[width=0.7\textwidth]{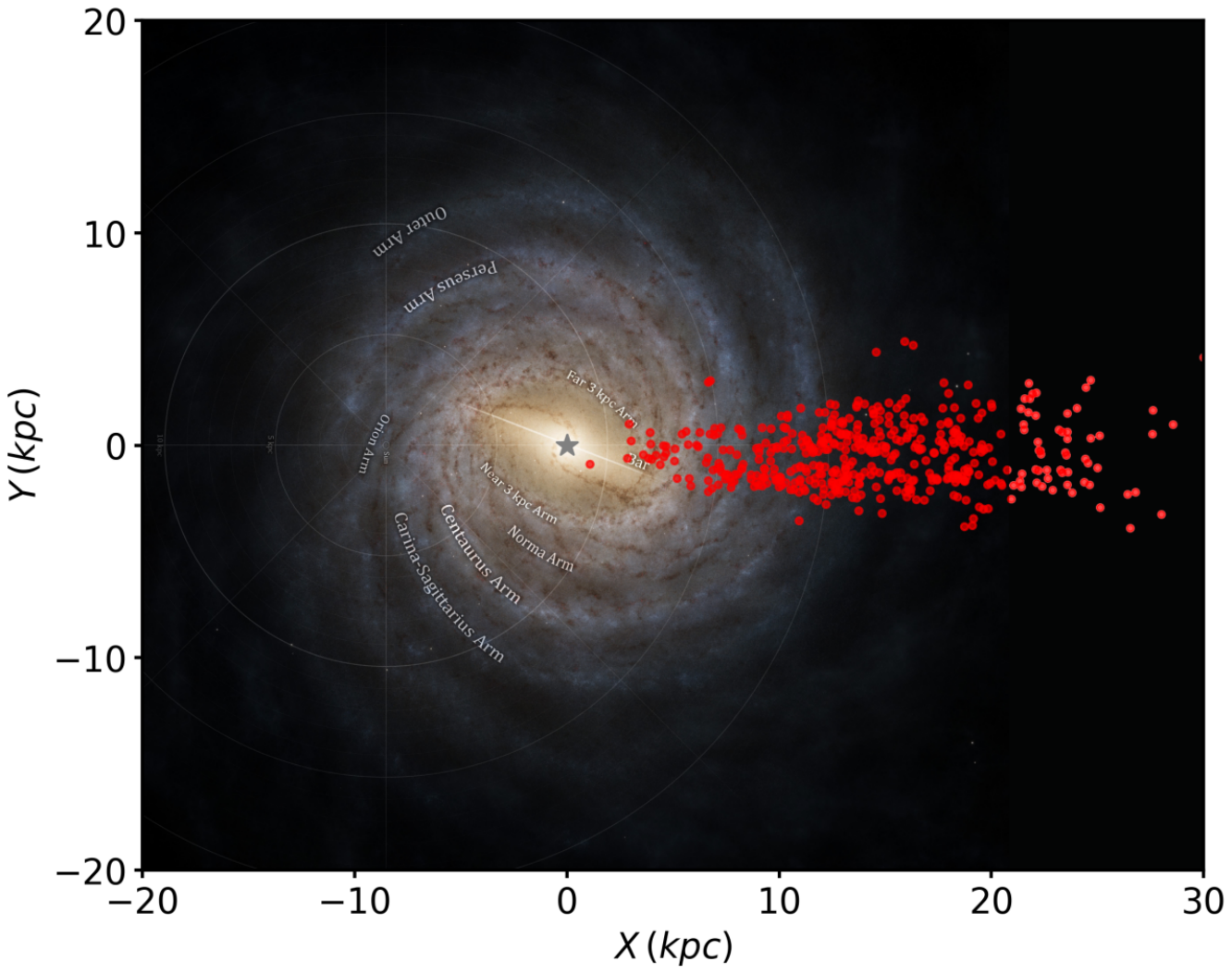}
\end{adjustwidth}
\caption{Distribution of {new, low-luminosity} RSGs identified out to 30 kpc using the granulation characteristics of OGLE light curves (reproduced from \citep{Zhang2025}).}\label{fig3}
\end{figure}

Systematic searches for RSGs in the Milky Way began with the onset of near-IR surveys. Massive clusters containing RSGs were identified by selecting groups of bright stars, leading to the discovery of several clusters rich in supergiants, e.g., \citep{Figer2006, Clark2009, Negueruela2010, Negueruela2011, Messineo2016}, located at the base of the Scutum--Crux arm near its interface with the Galactic Bar (see also \citep{Negueruela2012}). RSGs have been discovered in open clusters, e.g., the coeval cluster NGC 7419 \citep{Marco2013} and super star clusters, such as Westerlund 1 \citep{Clark2005} and the Galactic Center (GCIRS7 \citep{Blum1996}), as well as in the Perseus arm \citep{Dorda2018} and the Per OB1 association \citep{Humphreys1978, deBurgos2020}. \citet{Healy2024} performed a comprehensive study of RSGs in the Milky Way, which extended to 12.9 kpc and included 578 highly probable RSGs. This is only a small fraction of RSGs in the Milky Way. The total number was estimated to be $\sim$5000 by \citet{Gehrz1989}, but more recently was increased to over 11,000 RSG candidates by \citet{Zhang2024}. These candidates were identified by combining low-resolution \textit{Gaia} spectra with color criteria, and therefore the number is expected to be more accurate. \citet{Zhang2025} identified over 450 new, {low-luminosity} RSGs in the Milky Way out to 30 kpc {(see Figure~\ref{fig3})} by analyzing the granulation characteristics of OGLE light curves, in combination with \textit{Gaia} DR3 parameters to eliminate red giant branch stars.

\begin{figure}[H]
\begin{adjustwidth}{-\extralength}{-4cm}
\centering
\includegraphics[width=1.0\textwidth]{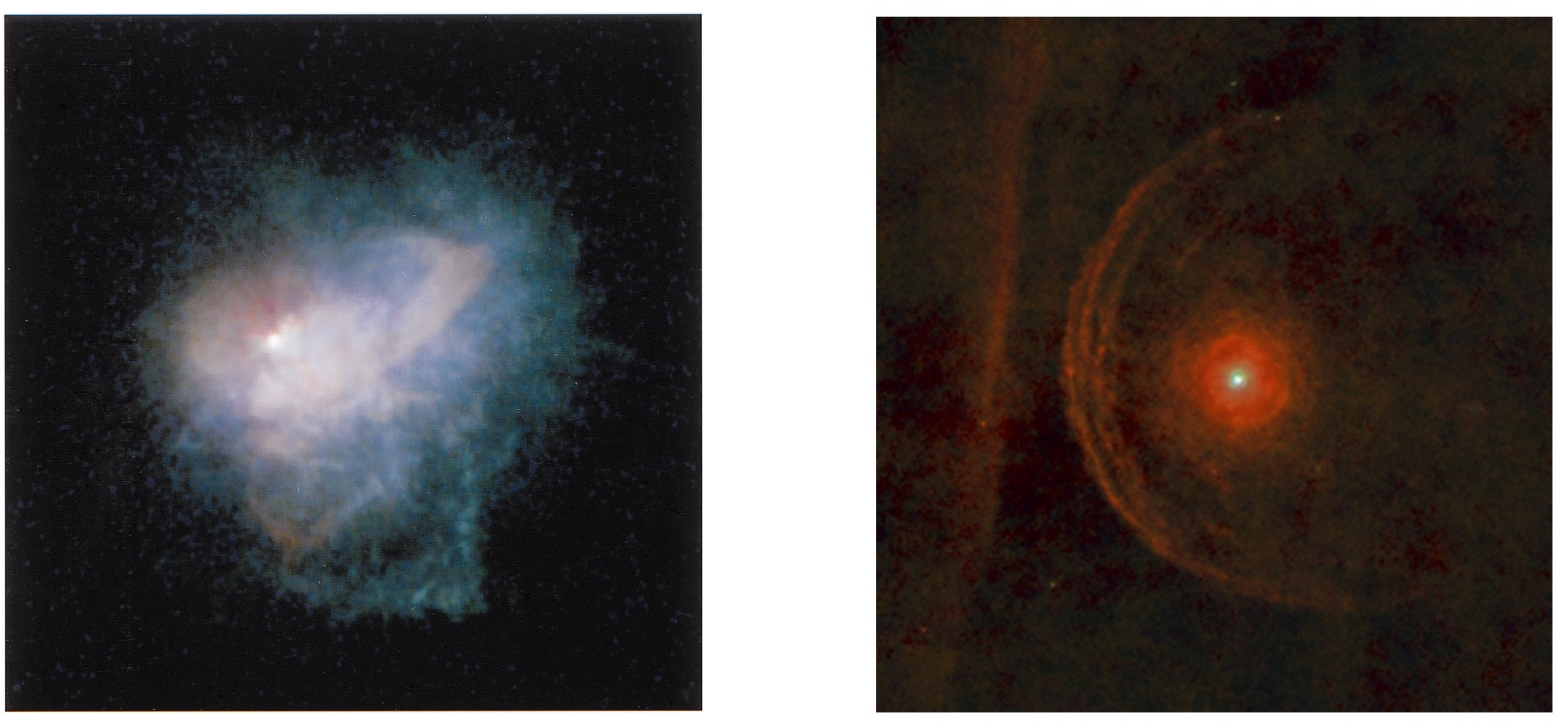}
\end{adjustwidth}
\caption{Examples of resolved RSGs in the Milky Way. \textbf{Left}: \textit{HST} composite image of VY CMa revealing the structure of its circumstellar material (reproduced from \citep{Smith2001}). \textbf{Right}: Bow shock and bar feature of Betelgeuse resolved with \textit{Herschel} [Image credit: ESA/\textit{Herschel}/PACS/MESS].}\label{fig4}
\end{figure}

The RSGs in our Galaxy offer us the opportunity for detailed studies of the circumstellar environment, and in particular of the mass-loss events that have occurred in their recent history. The nearest RSGs can be resolved with high-resolution imaging usin, e.g., \textit{HST}, e.g., VY CMa \citep{Smith2001}, NACO, and SPHERE on VLT, e.g., Betelgeuse \citep{Kervella2009,Montarges2021}, the MMT, and \textit{SOFIA}, e.g., \citep{Shenoy2016, Gordon2018} and interferometry, using, for example, GRAVITY or MATISSE on the VLT interferometer, e.g., \citep{Montarges2021,GonzalezTora2024}, 
ALMA, e.g., VY Canis Majoris \citep{Richards2014} or CHARA, e.g., $H$-band interferometry of AZ Cyg and RW Cep \citep{Norris2021, Anugu2023, Anugu2024}. Such high-resolution imaging observations have revealed complex circumstellar media, implying multiple asymmetric ejections, e.g., VY CMa \citep{Smith2001}; see Figure~\ref{fig4}. In addition, bow shocks have been discovered around several RSGs in the Milky Way: Betelgeuse \citep{NoriegaCrespo1997} (see Figure~\ref{fig4}), $\upmu$ Cephei \citep{Cox2012}, IRC $-$10414 
 \citep{Gvaramadze2014}, and W237 in Westerlund~1 with JWST \citep{Guarcello2025}. Evidence for a bow shock around the luminous RSG [W60] B90 in the LMC was presented by \citet{MunozSanchez2024}, making it the first extragalactic RSG with a bow shock and a massive analog of Betelgeuse.

\vspace{-6pt}
\unskip

Galactic RSGs have also been crucial in the determination of the RSG mass-loss rates. \citet{Beasor2020} used coeval RSGs in open clusters to derive their mass-loss rates and a steep mass-loss rate relation with luminosity, as a function of initial mass. They found much lower mass-loss rates than previous studies, implying that quiescent mass loss cannot strip the hydrogen envelopes of RSGs, therefore, affecting their evolution. \mbox{\citet{Decin2024}} resolved 5 RSGs with ALMA and measured independent, gas mass-loss rates based on CO(2--1). \citet{Antoniadis2024} managed to reconcile the dust and gas mass-loss rates, confirming the lower values found from these studies. This topic is reviewed in the chapter by van Loon in this Special Issue. Finally, the topic of RSG binaries in our Galaxy is reviewed in the introductory chapter by Humphreys.

\section{RSG Population of the Local Group}
\label{Section4}

The Local Group galaxies contain stars at a large range of metallicities, from \mbox{$Z=0.06$~Z$_{\odot}$} in Sextans A \citep{kniazev2005} and 0.14~Z$_{\odot}$ in WLM \citep{Urbaneja2008} to 1.5~Z$_{\odot}$ in M31 \citep{Zurita2012}. Moreover, the galaxies of the Local Group do not suffer from the distance uncertainties and high extinction that plague studies of the RSGs in the Milky Way. Distances to these galaxies are relatively well measured; for example, detached eclipsing binaries have provided a 1\% distance to the LMC \citep{Pietrzynski2019}, while the Cepheid period--luminosity relation  \citep{Leavitt1912} and the tip of the red giant branch method \citep{Lee1993} have provided uniform and accurate distances with typical statistical errors of $\sim$5\% (total errors including systematics are larger) for most of the galaxies in the Local Group (except for M33, see, e.g., \citep{Bonanos2006}) and beyond. Armed with accurate distances, the availability of accurate optical and, in particular, near-IR photometry from wide-field surveys to eliminate giants and \textit{Gaia} data to remove foreground sources, the RSG population of the galaxies in the Local Group has started to emerge.

As mentioned above, the wide-field \textit{UBVR} CCD surveys of the Magellanic Clouds by \citet{Massey2002}, the UBVRI survey of M31 and M33 by \citet{Massey2006}, and of seven dwarf galaxies---IC 10, NGC 6822, WLM, Sextans B, Sextans A, Pegasus, and Phoenix---by \mbox{\citet{Massey2007}} provided the first photometric catalogs for basing the selection of RSGs in the Local Group. The availability of multi-object slit or fiber spectrographs enabled follow-up spectroscopy for hundreds of objects in these galaxies. \citet{Massey1998a} and \citet{Massey2003} were the first to perform such follow-up surveys, confirming over $\sim$90\% of their optically-selected RSG candidates. These pioneering works were followed by other spectroscopic studies confirming photometric candidates in the Magellanic Clouds \citep{Levesque2006, deWit2023}, M31 \citep{Massey2016} and M33~\citep{Drout2012}, as well as multiple studies on dwarf irregular galaxies: WLM, NGC 3109, NGC 6822, Sextans A, IC 1613, Pegasus, Phoenix, and IC 10 
 \citep{Britavskiy2014, Britavskiy2015, Britavskiy2019, GonzalezTora2021, Bonanos2024, deWit2025}).

In the last 5 years or so, entire populations of RSGs in many Local Group galaxies have been identified and revised by improving the photometric selection criteria: \mbox{\citet{Yang2019}} identified 1405 RSGs in the SMC, while \citet{Yang2020} refined the catalog to 1239 RSGs, estimating the total number to be $\sim$1800. \citet{Yang2023} improved the catalog cleaning and completeness, identifying 2121 RSGs. In the LMC, \citet{Yang2021} found \mbox{2974 RSGs}. \citet{Antoniadis2024} refined the sample to 2219 RSGs. \citet{Ren2021b} reported 4823 and 2138~RSGs in the LMC and SMC, respectively. The latter study also identified RSGs in another 10 low-mass galaxies of the Local Group: WLM, IC 10, NGC 147, NGC 185, IC 1613, Leo A, Sextans B, Sextans A, NGC 6822, and Pegasus Dwarf. \mbox{\citet{Yang2021}} identified \mbox{$\sim$300 RSG} in NGC 6822 and compared the results of different selection methods. \mbox{\citet{Antoniadis2025}} have identified 82 more RSGs in NGC 6822 from the JWST photometry of \mbox{\citet{Nally2024}}. \citet{Boyer2024} used JWST photometry to separate RSGs from thermally-pulsing AGB stars in WLM. In M31 and M33, \citet{Massey2021} found {6412} RSGs in M31 and {2858} RSGs based on a near-IR photometric survey. \citet{Ren2021} independently identified 5498 RSGs in M31 and 3055 RSGs in M33 {using the same method, finding the estimated RSG populations for the two galaxies to agree within 10--15\%.}


\section{RSGs Beyond the Local Group}

Beyond the Local Group, several works have measured the parameters and metallicities of RSGs by obtaining $J$-band spectroscopy of targets in the Sculptor Group, e.g., NGC~300 and NGC~55 \citep{Gazak2015, Patrick2017}. The ASSESS project obtained low-resolution optical spectra of RSG candidates, producing the largest catalog of RSGs at low $Z$ beyond the Local Group (127~RSGs, analyzed by \citep{deWit2024}) in NGC 55, NGC 247, NGC 253, NGC 300, NGC 1313, M83, and NGC 7793 
 \citep{Bonanos2024, deWit2025}. This sample includes two spectroscopically confirmed RSGs beyond 4~Mpc, in M83 and NGC~1313. These are the most distant RSGs currently confirmed. Photometric RSG candidates in M83 have been identified with mid-IR colors by \citet{Williams2015}. Beyond that, super star clusters dominated by RSGs have been used as abundance probes \citep{Davies2010} in M83 \citep{Davies2017} and NGC~4038 in the Antennae \citep{Lardo2015}.

\citet{Chun2017} used near-IR photometry to identify RSGs in NGC 4449, NGC 5055 (M63), and NGC 5457 (M101), which extend from 4 to 9~Mpc. However, this study took place before the near-IR criteria to remove AGB stars \citep{Yang2019} and before \textit{Gaia} DR2; therefore, they could not perform foreground cleaning. The machine-learning algorithm of \mbox{\citet{Maravelias2022}} has identified over 120,000 robust candidate RSGs in nearby galaxies extending out to 5~Mpc, including over 80,000 candidates in M31, 25,000 in M33, $\sim$7000 
 in NGC~6822, 2300 in IC~1613, and over 600 in NGC~2403 and in IC~10 \citep{Maravelias2025}. Most of these are expected to be true RSGs, given the high recovery rate of the machine-learning algorithm for this class.

\section{Insights from Extragalactic RSGs}

Increased samples of extragalactic RSG with accurately measured parameters (T$_{\rm eff}$, L, R) have greatly impacted our understanding of their physical properties and evolution. For example, the dependence of T$_{\rm eff}$ and the Hayashi limit on metallicity \citep{Elias1985,Levesque2006, GonzalezTora2021} resulted from the spectroscopic follow-up of RSGs in the Magellanic Clouds. \citet{Davies2013} also found that effective temperatures of RSGs in the Magellanic Clouds derived from TiO bands were lower than those determined from SED fitting, because molecular bands are formed in the cooler, outer layers of the extended atmospheres of RSGs. \citet{Davies2021} further found that high mass-loss rates increase the strength of the TiO absorption bands, and that fluctuations in the wind can cause changes in spectral type. It remains unclear whether or not the extreme spectroscopic variability of Levesque--Massey variables \citep{Massey2007b, Levesque2007}, found primarily in low metallicity galaxies, is also caused by this mechanism. Disentangling the effects of metallicity, episodic mass loss, and T$_{\rm eff}$ requires improved, 3D, atmospheric models as the approximations made in the current ones cannot reproduce the complex atmospheres. Currently, this issue can be tackled with $J$-band spectroscopy and the modeling of atomic lines, e.g., \citep{Davies2015, Gazak2015, Patrick2015}, which yield both the metallicity and T$_{\rm eff}$. 

Extragalactic RSGs were also key to determining that RSG mass-loss rates are lower, e.g., \citep{Antoniadis2024, Antoniadis2025} than previously considered and implemented in evolutionary models (cf. \citep{Meynet2015, Zapartas2025}), that the mass-loss rate relation with luminosity has a `kink' around \mbox{$\log L/L_{\odot} \sim 4.5$ \citep{Yang2023, Humphreys2020}}, and that a significant fraction have dusty ejecta \citep{deWit2024, Antoniadis2025}. The chapter by van Loon in this Special Issue further discusses these results and their implications. Furthermore, extragalactic RSGs enabled the construction of luminosity functions, e.g., in M31 and M33, see \citep{Neugent2020, Massey2023}, to directly test evolutionary models and, in particular, the position of the SN progenitors; see, e.g., \citep{Meynet2015, Zapartas2025}. They also determined an upper limit of $\log L/L_{\odot} \sim 5.4$, {which represents the upper limit of RSG progenitors of type IIP supernovae}. Finally, extragalactic RSGs can be used to constrain the various physical processes taking place, such as the driving mechanism of the winds, episodic mass loss, and the extreme phenomena observed at high luminosities, as well as the binary fraction of RSGs. The following subsections describe some of these in detail.

\subsection{The Humphreys--Davidson Limit for RSGs}

Improved distances to nearby galaxies combined with high-quality photometric data demand a reassessment of the Humphreys--Davidson limit \citep{Humphreys1979}, which was originally reported to be at $\log L/L_{\odot} = 5.8 \pm 0.1$. This value is supported by more recent studies reporting high luminosity RSGs in M31 and M33, e.g., \citep{Gordon2016, Maravelias2025}; these sources deserve follow-up. However, \citet{Davies2018} found the upper luminosity limit of RSGs in both the Large and Small Magellanic Clouds to be $\log L/L_{\odot} \sim 5.5$. \citet{McDonald2022} determined an upper luminosity limit of $\log L/L_{\odot} \sim 5.53 \pm 0.03$ for RSGs in M31, finding that many sources with higher luminosity estimates turned out to be foreground stars, or were resolved as binaries or clusters. These works therefore hint at a lower, metallicity-independent value for the HD limit in the regime of RSGs. Repeating this exercise in more galaxies will help confirm this result. \citet{Martin2023} identified two RSGs in the LMC that are near the higher value limit: WOH G64 and IRAS 05346-6949. However, WOH G64 has extended circumstellar material that was resolved by \mbox{\citet{Ohnaka2024}}, who also estimated a more accurate luminosity for the star, by taking into account the CSM. This value was 0.3 dex lower than the original estimation (although differences in the modeling assumptions of the dust may affect the luminosity; see, e.g., \citep{Humphreys2020}). By analogy, IRAS 05346-6949 and other RSGs with $\log L/L_{\odot} \sim5.8$ may also be revised to lower luminosities once their CSM is resolved. \citet{MunozSanchez2024} pointed out that Var~A and other extended sources may be overestimated by 0.3 dex similarly to WOH~G64, which would indeed bring the HD limit for RSGs down to $\log L/L_{\odot} \sim 5.5$. 

\subsection{Extreme RSGs}


Extreme RSGs are defined as RSGs that are extreme in terms of both luminosity and size, pushing the limits of hydrostatic stability of the star. The evolution of these RSGs remains unclear, given the uncertainties in the mass-loss relations used by stellar evolution models in both the hot star phase and the RSG phase (see, e.g., \citep{Smith2014, Antoniadis2024}), the amount of stripping predicted for various mass-loss rate relations \citep{Zapartas2025}, the effect of binary interactions for RSGs with companions, and that of episodic mass loss, which is not yet incorporated in evolutionary models. In depth studies of individual extreme RSGs offer great insight into these rare objects (e.g., see VY CMa and NML Cyg, mentioned above). Such studies have been recently extended to extragalactic RSGs, such as WOH~G64 \citep{MunozSanchez2024b} in the LMC, which have the benefit of accurate distances and therefore accurate luminosities.

WOH~G64 is unique in that it has offered us the opportunity to observe post-RSG evolution in real time. It was one of the most luminous ($\log(L/L_{\odot}$) = 5.45 dex \citep{Ohnaka2008}) and highly mass-losing RSG in the LMC ($\dot{M}>10^{-4}\, \rm M_{\odot} \,yr^{-1}$ \citep{Beasor2022, Antoniadis2024}) until its transition to a YHG in 2014, in a symbiotic B[e] binary, discovered by \citet{MunozSanchez2024b}. It is also the first extragalactic star to be resolved with interferometry (GRAVITY on VLTI \citep{Ohnaka2024}). The nature of its transition, which led to the ejection of the outer envelope of the star, remains unclear. The prevailing scenarios at the moment are (a) a pre-supernova superwind phase caused strong pulsations that ejected the envelope, (b) grazing envelope interactions in the binary disrupted the outer envelope, and (c) WOH G64 was always a YHG, but appeared as as RSG after a high mass-loss event that created a cool pseudo-atmosphere, as in the case of Var A in M33 \citep{Humphreys2006}.

\subsection{Dimming Events and Other Photometric Variability}

The semi-regular or irregular photometric variability of RSGs is due to a combination of factors, such as radial pulsations, the effects of convection at the surface of the star, such as shocks, hot spots, and cell turnover, see, e.g., \citep{Levesque2017}, and in some cases, binarity and mass loss. Photometric time series of extragalactic RSGs revealed the period--luminosity relation for RSGs \citep{Yang2011, Yang2012, Chatys2019}. Furthermore, the long secondary periods (LSPs) observed in many RSGs were {proposed as} the effect of a binary companion \citep{Goldberg2024}.

\begin{figure}[H]
\begin{adjustwidth}{-\extralength}{-4cm}
\centering
\includegraphics[width=1.2\textwidth]{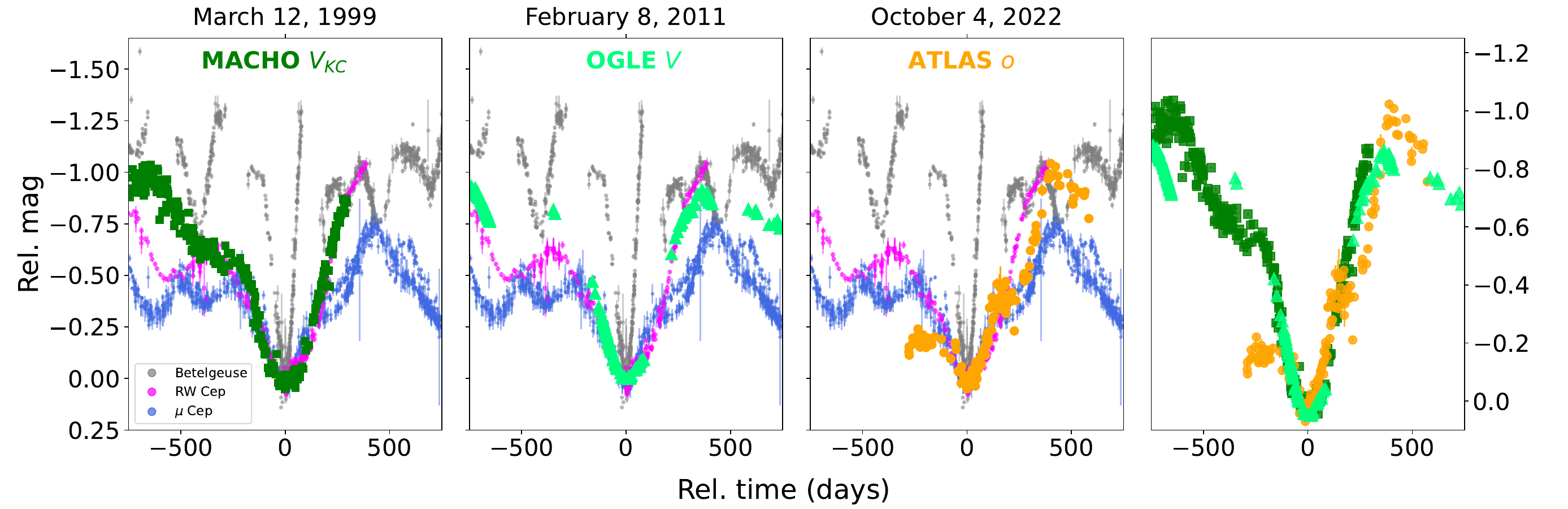}
\end{adjustwidth}
\caption{The three dimming 
 events in [W60] B90 ({dark and
bright green, and orange points, respectively} \citep{MunozSanchez2024}). The first three panels compare the dimming events {of [W60] B90} with the Great Dimming of Betelgeuse (gray), the dimming of RW Cep (magenta), and a minimum from $\upmu$ Cep (blue) in the $V$-band from AAVSO. The right panel overplots the three dimming event light curves of [W60] B90. The minimum of each dataset was used as the zero-point reference for each panel. (reproduced from \citep{MunozSanchez2024}).}\label{fig5}
\end{figure}

{Over the past 5 years, reports of dimming events in a handful of RSGs are offering great insight into the frequency and characteristics of episodic mass loss. The luminous RSG in the LMC, [W60] B90 ($\log L/L_{\odot} = 5.32$ dex \citep{deWit2023}) was} found to exhibit three dimming events with $\Delta$V$\sim$1~mag \citep{MunozSanchez2024} (see Figure~\ref{fig5}), similar in depth to the Great Dimming of Betelgeuse $\Delta$V$\sim$1.1~mag. However, these had a recurrence period of 11.8 years and a rise time of $\sim$400 days vs. $\sim$200 days for Betelgeuse. Dimming events with a similar depth of $\Delta$V$\sim$1~mag and rise time of $\sim$400 days were also reported in RW Cep \citep{Anugu2023} and $\upmu$ Cep (a minimum, in this case). Figure~\ref{fig5} compares the dimming events in the four stars mentioned above. The light curve rise time was found to correlate with the radius of the star, implying that smaller RSGs recover faster from dimming events, while larger RSGs take longer, which is likely related to the time needed for the atmosphere to readjust after a large convective cell rises to the surface and releases material. A calibration of this relation would make it a powerful distance indicator for RSGs, in particular those in the Milky Way, as this method is not affected by extinction. VY CMa has also exhibited a dimming event of \mbox{$\Delta$V$\sim$3~mag \citep{Humphreys2021}}, with a rise time of $\sim$500 days, in agreement with its larger radius. Another extreme dimming event ($>$2 mag in F814W) was reported by \citet{Jencson2022} in a \mbox{$\log L/L_{\odot}\sim$ 5.2 RSG} in M51, although it was not constrained well in time. 

Extending such studies to large samples of extragalactic RSGs will help determine the frequency and characteristics of dimming events as a function of luminosity.

\subsection{RSGs in Binary Systems}

The last few years have seen a renewed interest in RSGs in binary systems, following the finding that binary interactions dominate the evolution of massive stars and that over 70\% of all massive stars are expected to interact \citep{Sana2012}. Several studies have been undertaken, resulting in great leaps in our understanding of binary RSGs. \citet{Neugent2019} list the 11 Galactic binary RSGs known before 2019, all with spectroscopically confirmed B-type companions, as these have the largest contrast and are therefore the easiest to detect. RSGs in binary systems are expected to be in orbits with periods longer than 1000 days, so that the RSG does not fill its Roche lobe \citep{Sana2012, Neugent2020b}.

\textls[10]{The survey of cool supergiants in the Magellanic Clouds by \citet{GonzalezFernandez2015}} found about a dozen binary RSG candidates. It was followed by more systematic surveys for binary RSGs. \citet{Neugent2019} used photometric color-criteria to select UV-bright RSGs as candidate binaries and confirmed 87 binary RSGs spectroscopically, 63 in M31 and M33, and the rest in the Magellanic Clouds \citep{Neugent2018}. \citet{Neugent2019} increased the LMC sample by 38 binaries, while \citet{Neugent2020b} determined the binary frequency in the LMC to be $\sim$20\%. \citet{Neugent2021} found a higher binary frequency of $\sim$30\% in M31, and a clear metallicity-dependence in M33, with the frequency dropping from $\sim$40\% in the center to 16\% in the outer regions. \citet{Patrick2022} used photometry from the Ultraviolet Imaging Telescope (UVIT) on board \textit{AstroSat} to determine the intrinsic binarity fraction of RSGs in the SMC to be $\sim$19\%, while \citet{Patrick2024} confirmed 16 B-type companions of RSGs with ultra-violet spectroscopy. 

Two recent studies have presented evidence for a low-mass companion to Betelgeuse to explain its long secondary period \citep{Goldberg2024, MacLeod2025}. The mass of the companion is inferred to be $\lesssim 1-2$~M$_{\odot}$, which, if conatal, implies a pre-main sequence star. This opens a new parameter space for low-mass RSG companions, which at the moment elude detection. Finally, there have been searches for binaries with exotic companions among extragalactic RSGs, such as neutron stars that could appear as Thorne--$\rm \dot Z$ytkow objects. \citet{Levesque2014} reported the first such candidate in the SMC, while \citet{DeMarchi2021} suggest the detection of such exotic binaries with gravitational waves and multi-messenger observations.

\subsection{Red Stragglers}

Red stragglers are RSGs in a cluster that are more luminous and massive than expected from single-star evolution. They are thought to be stars rejuvenated by mass accretion and mergers in binary systems, and are therefore descendants or evolved products of blue stragglers. \citet{Britavskiy2019b} were the first to identify such stars and to coin the term `red stragglers'. They were found as luminous RSG outliers in two open clusters in the LMC and had a large age spread. \citet{Beasor2019} arrived at the same conclusions by studying two open clusters in the Milky Way and two in the LMC. \citet{Patrick2020} found a red straggler fraction of 50\% among RSGs in the SMC. \citet{Wang2025}, however, found that binary effects alone cannot account for the luminosity spread in open clusters.

\section{Future Outlook}

The recent advancements outlined above, including the improvement of selection criteria, flagging foreground sources and contaminants, and the development of machine-learning algorithms for identifying RSGs, along with spectroscopic follow-up efforts and the expansion of archival data, have resulted in constraints that demand accurate modeling. Full 3D models of stellar atmospheres, e.g., CO5BOLD \citep{Freytag2012, Ma2024}, are needed to resolve the inconsistencies encountered with the 1D MARCS models \citep{Gustafsson2008}, such as the T$_{\rm eff}$ measured using different parts of RSG spectra. They will also be able to make predictions about the variability in RSG spectra and light curves, including dimming events. Furthermore, improved evolutionary models for single stars, including the new mass-loss rate prescriptions and episodic mass loss are needed for {accurately placing} RSGs on the Hertzsprung--Russell diagram and for {predicting} the location of SN progenitors. The next step would be to incorporate accurate single-star physics into binary evolution modeling, e.g., see POSYDON~\citep{Fragos2023, Andrews2024}.

{Observationally, the spectroscopic confirmation of RSGs at 1~Mpc or beyond remains `expensive' as it requires significant observing time on 8--10 m class telescopes. The same holds for highly extincted RSGs in the Milky Way, which require near-IR spectroscopy. The multi-object capabilities of both optical and near-IR spectrographs accelerate the rate at which spectra can be collected; however, given that the fields of view of multi-object spectrographs are typically smaller than the sizes of most nearby galaxies, multiple pointings are required. The most distant RSGs to be confirmed spectroscopically are at $\sim$4.5~Mpc (in NGC 1313 and M83 \citep{Bonanos2024}). Besides securing telescope time, current observational challenges facing the confirmation of an RSG candidate include removing contaminants, which can be mitigated using astrometric data from \textit{Gaia}, the position on near-IR CMD, and time-series photometric data, when available.}

{The future is promising, with JWST and other new missions that will characterize full populations of RSGs in nearby galaxies; interferometry, which will resolve a larger number of nearby RSGs; the ELT, which will make possible very high-resolution and high-sensitivity observations;} and LSST, which will provide unprecedented time-series photometry and reveal the variability properties of RSGs in galaxies beyond 10~Mpc. Finally, machine-learning algorithms and AI are expected to exploit the exponential increase of astronomical data {from existing and future large-scale surveys} and significantly advance our knowledge of the final evolutionary stages of massive stars in the next decade.

\vspace{6pt}

\funding{This research received no external funding.}
 

\dataavailability{No new data were created or analyzed in this study. Data sharing is not applicable to this article. 
}

\conflictsofinterest{The author declares no conflicts of interest.} 

\abbreviations{Abbreviations}{
The following abbreviations are used in this manuscript:\\

\noindent 
\begin{tabular}{@{}ll}
2MASS & Two-Micron All Sky Survey \\
AAVSO & American Association of Variable Star Observers\\
AI & Artificial Intelligence \\
ALMA & Atacama Large Millimeter/submillimeter Array \\
CHARA & Center for High Angular Resolution Astronomy \\
CMD & Color--Magnitude Diagram \\
{ELT} & {Extremely Large Telescope} \\
HST & Hubble Space Telescope \\
IR & Infrared \\
{JWST} & {James Webb Space Telescope} \\
LMC & Large Magellanic Cloud \\
{LSST} & {Legacy Survey of Space and Time} \\
{MARCS} & {Model Atmospheres with a Radiative and Convective Scheme} \\
MERLIN & Multi-Element Radio Linked Interferometer Network \\
MMT & Multiple Mirror Telescope \\
OGLE & Optical Gravitational Lensing Experiment \\
{POSYDON} & {POpulation SYnthesis with Detailed binary-evolution simulatiONs} \\
RSG & Red Supergiant\\
SED & Spectral Energy Distribution \\
SMC & Small Magellanic Cloud \\
WISE & Wide-field Infrared Survey Explorer \\
VLA & Very Large Array \\
VLT & Very Large Telescope \\
YHG & Yellow Hypergiant \\
ZAMS & Zero-Age Main Sequence \\
\end{tabular}
}




\begin{adjustwidth}{-\extralength}{0cm}

\reftitle{References}


\bibliography{ref.bib}




%


\PublishersNote{}
\end{adjustwidth}
\end{document}